Title:

# Network analyses reveal negative link between changes in adipose tissue GDF15 and BMI during dietary induced weight loss

Authors:

Alyssa Imbert[1,2,3]*, Nathalie Vialaneix[3]*, Julien Marquis[4], Julie Vion[1,2], Aline Charpagne[5], Sylviane Metairon[5], Claire Laurens[1,2], Cedric Moro[1,2], Nathalie Boulet[1,6] Ondine Walter[5], Grégory Lefebvre[5], Jörg Hager[5], Dominique Langin[1,2,7,8], Wim HM Saris[9], Arne Astrup[10], Nathalie Viguerie[1,2,7]§, Armand Valsesia[5]§

* equal contribution, § joint supervision

Affiliations:

[1] Institut National de la Santé et de la Recherche Médicale (Inserm), UMR1297, Institute of Metabolic and Cardiovascular Diseases, Team Metabolic Disorders and Diabesity, 31400, Toulouse, France;

[2] University of Toulouse, UMR1297, Institute of Metabolic and Cardiovascular Diseases, Paul Sabatier University, 31400, Toulouse, France;

[3] INRAE, UR875 Mathématiques et Informatique Appliquées Toulouse, F-31326 Castanet-Tolosan, France ;

[4] Université de Lausanne, Genomic Technologies Facility, 1015, Lausanne, Switzerland

[5] Nestlé Institute of Health Sciences, Metabolic Health Department, 1015, Lausanne, Switzerland

[6] Institut National de la Santé et de la Recherche Médicale (Inserm), UMR1297, Institute of Metabolic and Cardiovascular Diseases, Team Adipose tissue, microbiota and cardiometabolic flexibility, 31400, Toulouse, France;

[7] Franco-Czech Laboratory for Clinical Research on Obesity, Third Faculty of Medicine, Prague and Paul Sabatier University, Toulouse, France;

[8] Toulouse University Hospitals, Laboratory of Clinical Biochemistry, 31000, Toulouse, France;

[9] Department of Human Biology, NUTRIM School of Nutrition and Translational Research in Metabolism, Maastricht University Medical Centre, Maastricht, The Netherlands;

[10] Department of Nutrition, Exercise and Sports, Faculty of Sciences, University of Copenhagen, Denmark.

Short Title:

Human adipose tissue GDF15 upregulation with diet

Correspondence : armand.valsesia@gmail.com; nathalie.viguerie@inserm.fr

Grants:

This work was supported by Inserm, Paul Sabatier University, the Innovative Medicines Initiative Joint Undertaking (grant agreement n° 115372), and the Commission of the European Communities (FP6-513946 DiOGenes).

Disclosure Summary: DL is a member of Institut Universitaire de France; JM, AC, SM, OW, GL, JH and AV are full-time employee at Nestlé; WHMS reports having received research support from several food companies such as Nestlé, DSM, Unilever, Nutrition et Santé and Danone as well as Pharmaceutical companies such as GSK, Novartis and Novo Nordisk; he is an unpaid scientific advisor for the International Life Science Institute, ILSI Europe. AA reports grants and personal fees from Global Dairy Platform, personal fees from McCain Foods, McDonald's, Arena Pharmaceuticals Inc., Basic Research, Dutch Beer Knowledge Institute, Netherlands, Gelesis, Novo Nordisk, Denmark, Orexigen Therapeutics Inc., S-Biotek, Denmark, Twinlab and Vivus Inc., grants from Arla Foods, Denmark, Danish Dairy Research Council and Nordea Foundation, Denmark, outside the submitted work, and royalties received for the book first published in Danish as 'Verdens Bedste Kur' (Politiken; Copenhagen, Denmark), and subsequently published in Dutch as 'Het beste dieet ter wereld' (Kosmos Uitgevers; Utrecht/Antwerpen, Netherlands), in Spanish as 'Plan DIOGENES para el control del peso. La dieta personalizada inteligente' (Editorial Evergrá ficas; Léon, Spain) and in English as 'World's Best Diet' (Penguin, Australia). The other authors have nothing to disclose.

Contributions:

We thank the Functional Biochemistry Facility for expert assistance with ELLA immunoassays.

AI and NVia designed and performed the network analyses; JM set up, optimized and supervised QuantSeq experiments, AC performed sequencing experiments, SM prepared and qc-ed RNA samples, JV performed RT-qPCR on full AT RNA samples and ThP1 cells, NB isolated AT cells and extracted RNA, CL performed RT-qPCR on RNA of AT isolated cells, CM analyzed GDF15 plasma data, OW setup and directed NIHS biobank, GL helped with bioinformatics analyses, WHMS and AA: designed the DiOGenes clinical study, AV, NVig, WHMS, JH, and DL: designed the transcriptomics studies, AV and NVig directed and supervised the whole study, AI, NVia, NVig and AV performed all statistical analyses, interpreted the results and wrote the manuscript with input from all authors; AV and NVig had primary responsibility for the final content; and all authors read and approved the final manuscript.

## Abstract

*Context.* Adipose tissue (AT) transcriptome studies provide holistic pictures of adaptation to weight and related bioclinical settings changes.

*Objective*. To implement AT gene expression profiling and investigate the link between changes in bioclinical parameters and AT gene expression during three steps of a two-phase dietary intervention (DI).

*Design.* AT transcriptome profiling was obtained from sequencing 1051 samples, corresponding to 556 distinct individuals enrolled in a weight loss intervention (8-week low calorie diet (LCD) at 800 kcal/d) followed with a 6-month *ad libitum* randomized DI.

*Methods.* Transcriptome profiles obtained with QuantSeq sequencing were benchmarked against Illumina RNAseq. RT-qPCR was used to further confirm associations. Cell specificity was assessed using freshly isolated cells and THP-1 cell line.

*Results.* During LCD, five modules were found, of which three included at least one bio-clinical variable. Change in BMI connected with changes in mRNA level of genes with inflammatory response signature. In this module, change in BMI was negatively associated to changes in expression of genes encoding secreted protein (*GDF15*, *CCL3* and *SPP*1).

Through all phases of the DI, change in *GDF15* was connected to changes in *SPP1*, *CCL3*, *LIPA* and *CD68*. Further characterization showed that these genes were specific to macrophages (with LIPA, CD68 and GDF15 expressed in anti-inflammatory macrophages) and *GDF15* also expressed in preadipocytes.

*Conclusion.* Network analyses identified a novel AT feature with *GDF15* upregulated with calorie restriction induced weight loss, concomitantly to macrophage markers. In AT, *GDF15* was expressed in preadipocytes and macrophages where it was a hallmark of anti-inflammatory cells.

## Precis

A data driven approach in obese humans during and after calorie restriction revealed adipose tissue *GDF15* gene expression upregulated during weight loss with a cluster of macrophages genes.

## Introduction

Combining biomarker and phenotypic data provides valuable opportunity to identify signatures of diseases as well as response to treatments or lifestyle interventions, which may have implications for understanding biology and clinical management (1–3).

Excess adiposity is associated to numerous comorbidities including metabolic complications such as insulin resistance (IR), type 2 diabetes (T2D), and cardiovascular diseases (CVD) (4,5). Adipose tissue (AT) is the main lipid storage organ of the human body. Through active secretory functions, it may directly, or indirectly via cognate receptor, influence the activity of other metabolic organs such as liver and skeletal muscle. AT secreted proteic factors, so-called adipocytokines are produced by either adipocytes, precursor cells or resident AT macrophages. These cell types modify gene expression in response to AT expansion or reduction. Furthermore, an AT dysfunction leads to an altered secretory profile and is associated with increased inflammation and fibrosis (6). The plasticity of AT during weight loss has been assessed by several gene expression studies, including transcriptome-wide analyses (7) and targeted candidate approach (8). Yet, changes occurring following weight loss and their link with weight regain still remain far from complete (9).

Additionally, very little is known about the relationship between weight regain, AT gene expression changes, and other clinical readouts such as indices of insulin resistance, and markers of CVD.  So far, most studies have focused on pairwise assessment between a given transcript and a single clinical readout. *De facto,* this limits our understanding of the physiological changes. Systems biology aims to dissect complex relationships across the multiple scales of organization that characterize biological systems (10,11). In particular, networks have proven useful to unravel the complex relations (regulation, co-regulation) existing between gene expression profiles under various environmental conditions (12). They are also a powerful approach to provide a global and comprehensive image of the systems functioning related to complex traits by studying jointly multiple clinical parameters (13,14).

In this study, we aim to characterize AT gene expression changes during a two-phase dietary intervention in overweight and obese subjects using unsupervised and hypothesis-free methods. This allowed us to identify modules of co-regulated genes related with clinical parameters pertaining to weight regain, insulin-resistance and risk factors for developing CVD.

## Material and Methods

### Ethics

All studies were performed according to the latest version of the Declaration of Helsinki. Local ethics committees approved all procedures that involved human participants and written informed consent was obtained from all participants.

### Randomized dietary intervention study design

The DiOGenes study (15) is a pan-European, multi-center, randomized controlled dietary intervention program (NCT00390637). A CONSORT flowchart describing the intervention is shown in Figure 1A. In this study, 938 overweight/obese, non-diabetic subjects followed a low-calorie diet (LCD) for eight weeks using a meal replacement product (Modifast 800kcal/d, Nutrition et Santé, France). Subjects achieving at least 8% of body weight loss were then included in a 6-month randomized dietary intervention (DI) and were assigned to one of five *ad libitum* maintenance diets, consisting in low protein/low glycaemic index, low protein /high glycaemic index, high protein/low glycaemic index, high protein/ high glycaemic index, and control according to current national dietary guidelines. Abdominal subcutaneous AT biopsies were obtained by needle aspiration, about 10 cm from the umbilicus, under local anesthesia after an overnight fast. Plasma and AT samples were stored at -80°C until processing. BMI, total plasma lipid levels, waist circumference and HOMA-IR were obtained at baseline, upon weight loss and study termination. Lipid levels and HOMA-IR were quantified following an overnight fast.

## DiOGenes transcriptome analyses

Total RNA was extracted from AT samples as previously described (8). RNA samples were then quantified with a fluorimetric method (Ribogreen, Thermo Fischer) and their integrity evaluated on a fragment analyzer (Advanced Analytical). Good quality RNA was available for 471 individuals at clinical investigation day (CID)1 (baseline), 330 at CID2 (at the end of the 8-w LCD) and 250 at CID3 (at the end of the 6-month DI) (Figure 1B). After sample randomization, 500 ng RNA was loaded on multi-well plates, dried by vacuum concentration, and resuspended into 5 µL nuclease free water. Sequencing libraries covering the 3'-end of messenger RNA were prepared using the QuantSeq 3' mRNA-Seq Library Prep Kit from Lexogen, strictly following the manufacturer's recommendations. The optimal number of PCR cycles (15 cycles) was empirically evaluated by quantitative PCR. Libraries were all quantified with a fluorimetric method (Picogreen, Thermo Fischer) and their size pattern evaluated on a fragment analyzer (Advanced Analytical). Libraries were pooled equimolar by 96 and clustered at a concentration of 9 pmol on 4 lanes of single read sequencing flow cells (Illumina). Sequencing was performed for 65 cycles on a HiSeq 2500 (Illumina) using the SBS v4 chemistry (Illumina).

After demultiplexing with bcl2fastq (standard parameters), sequencing reads were trimmed with BBDuk (BBTools version 35.85, Bushnell B., sourceforge.net/projects/bbmap/) using the parameters k=13, ktrim=r, forcetrimleft=11, mink=5, qtrim=t, trimq=10, minlength=20, rcomp=f, and providing the sequence of the QuantSeq adaptors. Mapping to the human genome (built GRCh38.p2) was performed with RNA STAR (16) (version 2.3.0e and using default parameters). Gene count was performed with HTSeq (17) (version 0.5.4p3, with the parameters mode=intersection-nonempty, stranded=yes, a= 10, and type=exon). The annotation file used was based on GENCODE (18) release 25 but was filtered for transcripts not classified as pseudogenes and with a transcript support level "1" (for high quality annotation) or "NA" (for single exon transcripts). Only seven samples did not reach the manufacturer's recommended criteria of 3M sequencing reads. Those samples were reprocessed.

PCA was performed on log2 transformed count data to identify possible outliers and batch effects. This allowed to remove 3 atypical samples from the analysis (see Supplementary Figure S1). A second PCA without these atypical samples allowed to identify a possible blood contamination in most samples from UK. Samples with HBB larger than 20% (50 samples) were also removed from subsequent analysis, together with a remaining outlier (see Supplementary Figure S2). In addition, one plate with 79 samples had a higher variability than the other plates. Normalization was not able to correct this plate-effect but differential analysis with and without the samples from this plate gave very reproducible results (see Supplementary Figure S3). So we decided to remove the samples from this plate from the analysis as well. This resulted in using 918 samples from 556 unique individuals in the remaining of the analyses, distributed at the different time steps, as shown in Supplementary Figure S4.

## Statistical analyses of QuantSeq data

Unless specified otherwise, all statistical analyses were performed using R (version 3.4)(19). The overall analysis strategy is summarized in Figure 2.

**Differential analysis** Pairwise time step analyses were performed to detect differentially expressed genes between two CID (*i.e.*, contrasts: CID1 vs CID2, CID1 vs CID3 and CID2 vs CID3). For each analysis, raw count data were normalized using the TMM approach (20) implemented in the R package edgeR (21). Then, differentially expressed genes between the two conditions were extracted using a Negative Binomial test with a fixed effect for the individual and a log ratio test. Multiple test correction was performed with Benjamini-Hochberg (BH) False Discovery Rate (FDR)(22) within each contrast and significance was set at FDR 5%. Genes whose expression was found equal to 0 in more than 25% of the samples were removed from the analysis, as were genes whose expression was too low in one condition, resulting in an impossible estimation of their fold change (FC). In addition, since high dimensional data ($n < p$) would cause estimation issues for network inference methods, we further restricted our list of differentially expressed genes to those having an absolute log fold-change (FC) greater than $\log_2(1.3)$.

**Integration of clinical and transcriptome data** Associations between clinical variables (BMI, total lipid levels, waist circumference and HOMA-IR) and gene expression were tested using linear mixed effect models. Changes in gene expression (CPM $\log_2$ FC), gender and age were modelled as fixed effects; the center was modelled as a random effect. Adjustment for multiple testing was performed using BH correction.

**Network inference and mining** Similarly to (11), we performed a multi-step network inference to obtain a comprehensive model of the overall interactions between gene expressions and clinical variables. In such a model, nodes of the network correspond to genes or clinical variables and edges correspond to strong and direct interactions between changes in gene expressions and/or clinical variables between two-time steps.

Edges between genes and clinical variables were inferred using the aforementioned mixed effect models. Edges between genes were inferred using the graphical Lasso (GLasso (23) as implemented in the R package huge) on the logFC expression of differentially expressed genes for each contrast. Unlike pairwise measure of associations, such as Pearson correlation coefficients, GLasso is based on partial correlations and provides a stronger criterion for dependency by adjusting for common co-expressed genes. This method is useful in order to filter out false positives by discovering only the most direct interactions. Tuning of the GLasso regularization parameter was performed using the RIC criterion (see Supplementary Methods). Finally, an unsupervised clustering of the nodes was performed for the three networks using the modularity (24) optimization method of Reichardt et al. (25), as implemented in the R package igraph (26). This led to obtain strongly connected groups of genes and/or clinical variables for each network.

**Pathway analysis** The biological functions represented by genes in each module were searched using Ingenuity Pathways Analysis (IPA) software version 7.5 (Ingenuity Systems, Redwood City, CA). Genes for which IPA reported location as "Extracellular Space" were considered to encode secreted factors. The significance of canonical pathways was tested using the Fisher Exact test with the User Dataset of the 3 reference sets for each contrast.

## AT cell isolation

The AT fractionation was performed as described in (27). Briefly, after collagenase digestion (250 U/mL in phosphate-buffered saline (PBS), 2% bovine serum albumin (BSA), pH 7.4, volume/volume) of the AT for 30min at 37°C, the cell suspension was filtered through a 250 µm filter. The floating mature adipocytes were collected, washed 3 times and stored at -80 °C. The remaining stroma vascular fraction (SVF) was obtained after centrifugation. SVF cells were treated with erythrocyte lysis buffer (155 mmol/L $NH_4Cl$; 5.7 mmol/L $K_2HPO_4$; 0.1 mmol/L EDTA; pH 7.3) followed by successive filtrations through 100, 70, and 40 -µm strainers. The viable recovered cells were counted and, after washing, the different SVF cells were isolated using an immunoselection/depletion approach utilizing magnetic microbeads coupled to specific CD antibodies (CD31, CD34, CD14) which are membrane cell markers to select the different SVF cell types. The preadipocytes are CD34-positive cells (CD34+) and CD31-negative (CD31-) cells. The CD34-negative cells (CD34-) are immune cells (macrophages which are also CD14+ cells and lymphocytes which are CD14- cells) as described in (28). The cell extracts were collected and stored at -80°C until RNA extraction.

## THP-1 cell culture

THP-1 cells were used as a human macrophage cell model. The cells were cultured in a humidified incubator at 37 °C with 5% $CO_2$ in RPMI 1640 (Gibco) supplemented with 10% heat-inactivated FCS (VWR) and 100 units/mL penicillin, 100 µg/mL streptomycin and 10mM HEPES (Gibco). Cells were seeded at $5.5 \times 10^5$ cells/mL in 1 mL into 12 wells culture plates, then differentiated into M0 macrophage-like cells by stimulation with PMA 1 ng/mL (Sigma Aldrich) for four days followed by 48h without PMA. To alter the phenotype, macrophages were primed for 48h with fresh medium supplemented with LPS (2 ng/mL; Miltenyi Biotec) and IFN-γ (10 ng/mL; Miltenyi Biotec) to differentiate into the M1-like phenotype, or IL-4 (20 ng/mL; Miltenyi Biotec) to the M2-like phenotype. After washing, cell extracts were collected and stored at -80°C until RNA extraction.

## Adipose tissue explants

AT explants were used for ex vivo studies. AT samples of about 400 mg obtained from needle biopsies in 11 overweight (mean BMI 27.7 kg/$m^2$, SD 3.7) women aged 36.1 Y (SD 5.1) were cut into small pieces and incubated for 4h in 4ml of Krebs/Ringer phosphate buffer supplemented with 1g/L glucose and 20g/L BSA as described in (29).

## RT-qPCR validation

Three hundred fifty-nine (359) individuals had good quality RNA at all 3 CIDs. The cDNA prepared from 500 ng of total RNA and processed using the using Superscript II reverse transcriptase (Invitrogen, St Aubin, France) in the presence of random hexamers were analyzed using the StepOne Plus Real-Time PCR system (Applied Biosystems, Carlsbad, CA) and TaqMan assays (Applied Biosystems) as described in Sramkova *et al* (29). The Taqman assays were obtained from Applied Biosystems and the respective IDs were: *GDF15* (Hs00171132_m1), *SPP1* (Hs00959010_m1), *CD68* (Hs00154355_m1), *LIPA* (Hs01548815_m1), *CCL3* (Hs00234142_m1), *PSMC4* (Hs00197826_m1), *PUM1* (Hs01120030_m1), and *18S* (Hs99999901_s1). The relative gene expression was calculated as $2^{-\Delta Ct}$ using *PUM1* as reference gene for full AT samples, *PSMC4* for THP1 cells or *18S* for AT isolated cells data.

## ELISA assays

GDF15 concentration in the buffer used for explants experiments was assessed using the ELLA SinglePlex assays (ProteinSimple, San Jose, California, USA). Plasma protein levels of GDF15 were measured in duplicate using human GDF15 ELISA kit (Human GDF-15 Quantikine ELISA Kit, Bio-Techne), following manufacturer’s instructions. The GDF15 concentrations were calculated using sigmoidal standard curve fitted by nonlinear regression analysis for each test.

### Statistical analyses of RT-qPCR and ELISA data

Gaussian distribution was tested using the D'Agostino & Pearson test. As data normality was not fulfilled, comparison of expression distribution between groups was performed using Kruskall-Wallis or Friedman test for unpaired or paired data respectively and Dunn's multiple comparison test. Linear regression was performed with change in BMI as dependent variable.

## Results

### Baseline characteristics and overall dietary intervention outcome

In this report, we investigated gene expression changes within a subset of both men and women that had followed a two-phase dietary intervention (see Methods and Figure 1A) and for which RNA samples from AT biopsies were available (Figure 1B). At baseline, subjects were on average 41 years old, with a mean BMI close to 35 kg/m$^2$ and a mean HOMA-IR at 2.93 (Table 1). Upon LCD, individuals achieved, on average, 11% weight loss and upon study termination (6-months after weight loss) 10.8% weight loss. Both genders achieved similar weight loss relative to baseline (p=0.2376). These characteristics are representative of all 918 enrolled subjects in the DiOGenes study.

### Differential gene expression analysis

Upon quality control of the sequencing data (see Methods and Figure 1B), we assessed which genes were differentially expressed between each paired clinical intervention time points (CID).

After filtering of low-quality genes, 6,290 genes were found differentially expressed for the contrast CID1/2, 5,263 for the contrast CID1/3 and 4,461 for the contrast CID2/3. This resulted in a list of 9,156 unique genes that were found differentially expressed for at least one of the contrasts, among which 1,228 were found differentially expressed for the three contrasts, as shown in Supplementary

Figure S5. Among these genes, 541 had an absolute log FC > $\log_2(1.3)$ for the contrast for CID1/2, 470 for the contrast CID1/3 and 661 for the contrast CID2/3 (for a total of 1,160 unique genes, used for network inference as shown in Supplementary Figure S5). The list of expressed genes with results on filtering and differential analysis can be found as supplementary material (Supplementary File S1).

### Replication with RNASeq data

Taking advantage of previously generated RNAseq data for the CID1/2 contrast for 191 subjects (7), we compared the RNAseq and QuantSeq technologies. Based on expression levels from 19,938 genes, we found a very strong correlation in expression fold-changes during LCD (Pearson $r^2$=70% with 95% CI [69.7%-71.2%], see Supplementary Figure S6A). Next, we attempted to replicate the 541 genes found differentially expressed with the QuantSeq analyses during LCD. At FDR 5%, 481 of those genes replicated with RNAseq analysis (Supplementary Figure S6B), thereby demonstrating a 90% replication rate between the two technologies.

### Network analyses reveal GDF15 as a novel partner in adipose tissue inflammation related genes

We used a system biology approach to investigate the link between AT gene expression and clinical parameters during a 2-phase DI including a LCD and the subsequent 6-month weight follow-up.

The first network analysis investigated the link between changes in AT gene expression and clinical parameters during the LCD-induced weight loss phase (see Figure 3). The network was centered on a clinical parameter, BMI, with both positive and negative relationship to a bunch of genes. Clustering of the nodes (representing genes or clinical parameters with significant changes) of the graph was performed. It revealed 5 modules with respectively 111, 89, 41, 131, and 131 nodes, with 3 of these modules including at least one clinical parameter (See Supplementary File S2 for a full description of the modules). Most of the modules contained more than 70% of down-regulated genes, except for the module containing BMI that exhibited 89% of up-regulated genes. The lists of genes associated to clinical variables in each module are displayed in the Supplementary File S3.

Pathway analyses of the genes were therefore performed for each module (supplementary Table S1). Total cholesterol, LDL-cholesterol and HOMA-IR were included in the same module which contained 108 genes. “Cell development” and “lipid metabolism” were the top biological functions and pathways represented. The changes in the mRNA level for 28 genes were positively connected to changes in total cholesterol (and 6 genes with negative association). The change in LDL-cholesterol was connected to change in mRNA level for 3 genes: *SSTR2* (positively), *U1* and *U1.1* (negatively). These 3 genes were also connected to total cholesterol in the same manner. The lincRNA *RP3.483K16.4* was the single gene connected, positively, to HOMA-IR. The module including waist circumference contained 130 genes with 85% positively connected to waist circumference, indicating that the highest was the reduction in waist circumference the greatest was the downregulation of expression of these genes. “Lipid metabolism” was the top biological function represented. Four genes encoded secreted proteins (*APOC4*, *EGFL6*, *SEMA3C*, *TNMD*). The BMI centered module included 130 genes among which 80 were connected to BMI and, for 91%, with negative relationship (Supplementary Figure S7A). “Inflammatory response” was the top biological function represented with 51 genes. Fourteen genes encoded secreted factors (*CCL3, CHI3L1, CHIT1, FCGBP, FCGR3A, FCMR, GDF15, IFI30, PILRA, PLA2G7, SDC4, SPP1, TREM2, ZG16B*), among which 12 were negatively connected to BMI indicating that the highest was the decrease in BMI, the greatest was the up-regulation of expression of these genes.

Our two subsequent network analyses investigated changes during the weight follow-up (CID2/3) and the overall DI (CID1/3). We identified 7 modules (with size 20-83 nodes) for CID1/3 and 8 modules (size 6-142) for CID2/3 (see details in Supplementary File S2). Both networks exhibited a module containing BMI as biological variable. For each contrast, the biological functions represented in the modules with BMI were “organismal injuries” and “lipid metabolism” (supplementary Table

S1). The “lipid metabolism” pathway was representative of the module containing HDL for CID2/3. Regarding these two contrasts (CID1/3 and CID2/3), the modules containing the BMI were composed of genes representing different biological functions (lipid metabolism and organismal injuries, respectively). Also, both networks contained a module with a cluster of co-regulated genes composed of *CCL3, CD68, GDF15, LIPA,* and *SPP1* (Supplementary Figures S7B and S7C).

The cluster was also found in the module containing BMI from the LCD-induced weight loss phase (Supplementary Figure S7A). This observation led us to focus on *GDF15* which was recently reported as a nutritional stress marker (30), and the four other genes of the cluster (*CCL3, CD68, LIPA,* and *SPP1*) which are markers for AT macrophages (31).

## Validation using RT-qPCR

Validation of findings used RT-qPCR and a larger subset of the DiOGenes study (590 individuals, 116 men and 474 women at baseline, including 352 individuals with paired samples regarding all contrasts).

The negative association between the change in AT *GDF15* gene expression and the change in BMI was confirmed during LCD (contrast CID1/2). A low positive association was found during weight follow-up, but not across the whole dietary intervention (Supplementary Figure S8).

All five genes of the cluster of macrophage markers had relative expression significantly different between all pairwise time-contrasts (and with FDR adjusted p-values < 0.05). Notably all genes marked a strong up-regulation during LCD, followed with a down-regulation during the following 6-month weight control phase and with expression levels significantly lower than their baseline levels (Figure 4).

Reports of GDF15 in AT are scarce, and none describe the cell type of origin. Therefore, we compared *GDF15* expression between isolated adipocytes and stroma vascular cells isolated from human AT together with gene expression of *CCL3, CD68, LIPA,* and *SPP1*.

Compared to isolated adipocytes, *GDF15* mRNA levels were higher in the stromal fraction, mainly in pre-adipocytes and macrophages (Figure 5). The four other genes were predominantly expressed in macrophages.

We next investigated the expression profile of these 5 genes according to phenotypic profile of the human macrophages cell line THP-1. Once induced to M0 macrophages, the THP-1 cells were polarized to M1-like (pro-inflammatory) or M2-like (anti-inflammatory) phenotype. All genes exhibited differential expression between M1 and M2 THP-1 cells, with *CD68, GDF15,* and *LIPA* having higher expression in M2 than M1 macrophages, while *CCL3* and *SPP1* mRNA levels were higher in M1 cells (Figure 6).

## Analyses of GDF15 protein

Next, we sought to confirm the observed AT *GDF15* mRNA levels at the protein levels in AT and blood. The secretion of GDF15 by AT was assessed using subcutaneous AT explants from 11 overweight women. GDF15 concentration in the media from the AT explants was 440.9 ± 114.2 pg/mL (mean ± SD, data not shown). Changes in circulating GDF15 was investigated in a subset of 28 individuals with both plasma samples, clinical and AT RT-qPCR data available at all CIDs. At baseline plasma GDF15 was 473.5 ± 184.9 pg/mL. No significant change in plasma GDF15 was found across all phases of the dietary intervention (p= 0.4908, Supplementary Figure S9A). A positive association between plasma GDF15 and *GDF15* gene expression in adipose tissue was observed at baseline (Supplementary Figure S9B), but no correlation between change in AT *GDF15* expression and variations in plasma concentrations was found (p-value > 0.5). The negative association between change in plasma GDF15 and change in BMI during the LCD was previously reported (32), but no significant association was found during weight follow-up or across the whole DI (p-value > 0.4) (data not shown).

## Discussion

In this study, we investigated gene expression changes in AT, during a two-phase dietary intervention in overweight/obese, non-diabetic subjects. Our systems biology approach enables to relate such changes with changes in different clinical parameters associated to obesity or comorbidities (BMI, waist circumference, total lipid levels, and HOMA-IR). In a discovery phase, we implemented the QuantSeq technology (33) that enables to substantially reduce the sequencing costs (by about 5x only in term of reagent costs), while keeping high-quality transcriptomic profiling for all Human protein-coding genes. This allowed us to assess a large RNA collection (>1,000 samples stemming from a clinical intervention). Re-analysis with nearly 400 samples using Illumina RNASeq demonstrated a 90% replication rate. Also, our validation phase, using RT-qPCR in >1,000 samples further reproduced our initial findings, thereby validating the QuantSeq technology.

Through a network-based approach, we identified BMI as a key node, indicating that this specific clinical outcome has a major link with gene expression. Also, there was one specific transcriptomic cluster including the *GDF15* gene, whose expression change during weight loss was linked with BMI change. Of note, *GDF15* was not identified in previous differential gene expression analyses following LCD-induced weight loss (7). GDF15 (growth differentiation factor 15)/MIC-1 (macrophage inhibitory cytokine-1)/NAG-1 (nonsteroidal anti-inflammatory drug-activated gene) is a member of the transforming growth factor β (TGF-β) superfamily, that was first identified as a blocker of macrophage activation (34). Its expression is ubiquitous and circulating concentration range is high (35). Recently, GDF15 has generated considerable attention in the field of obesity and weight control. Notably, targeting GDF15 for the treatment of obesity and anorexia is the subject of several studies (35–40).

Preclinical studies in mice showed that GDF15 suppresses food intake (41) and recombinant GDF15 administration lowers body weight (42). Also, GDF15 can directly increase thermogenesis and improve insulin sensitivity (43). In human, a rise in blood levels was reported with acute exercise (32,44) and exercise training (45). Regarding weight loss, during or after termination of calorie restriction, an increase was observed in obese individuals upon bariatric surgery (46), and 2 weeks metformin-induced weight loss (47). Small-scale dietary studies also showed no (PMID: 32020057 DOI: 10.1038/s41430-020-0568-9) or slight increase in plasma GDF15 following light or drastic calorie restriction, for 48h or over 28 days (30). This was also confirmed in serum upon very-low calorie diet (48). In addition, T2D patients exhibited lower GDF15 plasma levels 6 months after the termination of an 8-week very-low calorie diet (PMID: 33925808 PMCID: PMC8146720 DOI: 10.3390/nu13051465). Hence, circulating GDF15 is considered as a nutritional stress marker (30).

Only few studies investigated the ATs (51,52). In our study, we found that GDF15 is released by AT and *GDF15* gene expression in AT was up-regulated upon an 8-week low calorie diet, in association with change in BMI and independently of change in plasma lipid profile or insulin resistance. The baseline plasma level was in range with other studies (37). No significant change was found in plasma ($p>0.4$). This might be due to the low power of the plasma study (only 28 individuals were investigated), or a low contribution of AT to circulating GDF15 compared to other tissues, as suggested by the lack of association between changes in AT *GDF15* expression and variations in circulating GDF15. In AT, we noticed that the *GDF15* up-regulation was only transient, as *GDF15* gene expression levels were significantly reduced in the 6-month following the acute weight loss phase despite sustained weight loss, compared to baseline. Importantly, AT *GDF15* mRNA levels were significantly reduced at study termination compared to baseline. However, plasma GDF15 remained steady. This indicates that AT *GDF15* upregulation is induced by a negative energy balance. Also, it tends to suggest that this is linked to AT remodeling during LCD, and that upon the acute weight loss phase, the metabolic improvements (both in term of weight loss and insulin sensitivity) relates to generally higher GDF15 levels. It is to be noticed that GDF15 induces lipolysis as recently reported (32).

The identification of GDF15 as a factor released by the AT led to the observation that this adipokine is produced by both adipocytes and stromal cells, with higher *GDF15* expression in subcutaneous than visceral AT and expression negatively associated with body fat mass in both fat depots (51). In human AT, *GDF15* expression was reported as a marker of oxidative stress negatively associated to lipogenic gene markers (52). In the present study, changes in *GDF15* mRNA levels are positively associated with changes in expression of macrophages markers (*CCL3*, *CD68*, *LIPA*, *SPP1*) in all contrasts. We show that *GDF15* expression arises from two human AT cell types, preadipocytes and macrophages, while expression of the four genes (*CCL3, CD68, LIPA* and *SPP1*), all co-expressed with *GDF15*, only displayed macrophages-specific profiles. Macrophages exhibit phenotypic heterogeneity and plasticity, depending on their microenvironment. These cells originate from circulating monocytes and infiltrate tissues where they play various functions including tissue cleanup and repair. The M1 and M2 classification is an oversimplification of a continuum in activation states, and individual markers may fail to specify such polarization phenotype (53). M1-like macrophages promote AT inflammation and insulin resistance; while M2 macrophages have an anti-inflammatory role (54). We found that *GDF15* was more expressed in M2-like macrophages. This is consistent with a mice study showing that *GDF15* expression was previously reported suppressed in M1-like macrophages (55). GDF15 also enhances the oxidative function of macrophages, leading to polarization into an M2-like phenotype (55). Interestingly our data show that preadipocytes which are part of the SVF cells, and not differentiated adipocytes, also express *GDF15* at similar level than macrophages. While preadipocytes *GDF15* expression studies remain scarce (52,56), these cells can be reprogrammed through dietary-induced weight loss and contribute to improvement of the metabolic syndrome (57). Further studies are needed to elucidate the potential contribution of preadipocytes to weight loss and related metabolic dysfunction improvements through *GDF15* expression.

We also report different expression profiles of the macrophage markers associated to *GDF15*. *SPP1* and *CCL3* expression levels were higher in M1-like macrophages. Osteopontin, encoded by *SPP1*, is an important component of immune response and inflammation (58). *SPP1* expression is positively associated to AT macrophages accumulation (59), and osteopontin plays a role in the development of insulin resistance (60). *CCL3* encodes MIP-1α, a member of the CC chemokine family that is produced by a variety of cells, including resident and recruited macrophages (61). Conversely, *LIPA* and *CD68* expression levels were higher in M2-like macrophages. *LIPA,* encodes the lysosomal acid lipase protein that breaks down cholesteryl esters and triglycerides in human macrophages. Its expression and activity have been reported to be decreased in the metabolic syndrome (62). *CD68* encodes a membrane protein marker and in our analyses, we observed its expression in both M1 and M2 macrophages (with slightly higher levels in the latter population). This is consistent with previous reports on CD68, that document it as a general marker of macrophages, whose expression is directly linked with the number of macrophages, and that associates with both pro- and anti-inflammatory markers (63).

In obese individuals, weight loss induces a decrease in pro-inflammatory and an increase in anti-inflammatory factors (64). The up-regulation of these five genes during LCD may be a hallmark of the beneficial effect of calorie restriction-induced weight loss on AT inflammation. Lending support to this hypothesis, a recent study showed that treatment of obese mice with GDF15 improves the oxidative function of AT macrophages and reverses insulin resistance (55). As no significant change in circulating GDF15 was found, the present study indicates a paracrine/autocrine role of GDF15 within AT. The study cannot provide evidence on whether the enhanced GDF15 expression during LCD originates from preadipocytes or macrophages, however, a strong co-regulation of both M1 and M2 macrophages markers was found. GDF15 appears as an anti-inflammatory marker. In addition to the decrease in the anti-lipolytic insulin, the pro-lipolytic GDF15 locally produced within AT may contribute to LCD-induced weight loss (32). The fatty acids produced by adipocytes could thereby induce transient local inflammation.

In summary, we identified an AT signature as a cluster of macrophage-related genes, through a transcriptome-wide systems biology approach. Specifically, a module including *GDF15* was identified; while *GDF15* is currently the focus of targeted studies, it demonstrates the validity of our approach to identify potentially relevant biomarkers of clinical improvements during dietary intervention. And indeed, our approach highlighted a novel macrophage signature composed of genes co-regulated with *GDF15*.

## Legends for Figures

Figure 1: Flowcharts of dietary intervention and expression studies. (A) Flowchart of the DiOGenes dietary intervention. (B) Flowchart of the DiOGenes gene expression analyses (path 1: discovery analyses with the use of RNAseq; path 2: validation analyses with the use of RT-qPCR). CID, clinical investigation day; DiOGenes, Diet, Obesity and Genes; LCD, low-calorie diet; QC, quality control; RT-qPCR, reverse transcription quantitative polymerase chain reaction; RNAseq, RNA sequencing.

Figure 2: Workflow of the samples, data and network analysis. CID, clinical investigation day; DEG, differentially expressed genes; FC, fold change; GMM, Graphical Gaussian Model.

Figure 3: Global network for changes in gene expression and bioclinical parameters during low calorie diet. A sparse Graphical Gaussian Model was used to estimate partial correlations in each set of variables (changes in adipose tissue mRNA level, and changes in bio-clinical parameters during LCD) and mixed models were used to assess links between gene expressions and bio-clinical variables. Network was laid out using force-based algorithms in Gephi 0.9.2 software. The bio-clinical variables are displayed with high size labels. Edge color indicates the correlation sign: red for positive correlations and blue for the negative ones. BMI, body mass index; HDL-Chol, high density lipoprotein; LDL-Chol, low density lipoprotein cholesterol; HOMA-IR, homeostatic model assessment of insulin resistance.

Figure 4: Adipose tissue gene expression using RT-qPCR at all time-points of the dietary intervention. The mRNA levels of *CCL3*, *CD68*, *LIPA*, *GDF15* and *SPP1* were measured in abdominal subcutaneous adipose tissue (n=219-351) at baseline (CID1), after an 8-week low calorie diet (CID2), and after 6

months of weight maintenance diet (CID3). CID, clinical investigation day. ***, p <0.001 from Friedman and Dunn's multiple comparison tests.

Figure 5: Adipose tissue gene expression in human adipose tissue cells. The mRNA level of *CCL3*, *CD68*, *LIPA*, *GDF15* and *SPP1* were measured in abdominal subcutaneous adipose tissue freshly isolated adipocytes (Adipo), preadipocytes (Preadipo), lymphocytes (Lympho) and macrophages (Macro)(n=5). *, p <0.05 ; **, p <0.01; ***, p <0.001 from Kruskal-Wallis and Dunn's multiple comparison vs. adipocytes tests.

Figure 6: Adipose tissue gene expression in pro- and anti-inflammatory macrophages. The THP-1 cell were induced to M0 macrophages, then polarized to M1-like (pro-inflammatory) or M2-like (anti-inflammatory) phenotype. The mRNA levels of *CCL3*, *CD68*, *LIPA*, *GDF15* and *SPP1* were measured in M0, M1 and M2 macrophages. The data are presented normalized to M0 phenotype (n=11). *, p <0.05 ; **, p <0.01; ***, p <0.001; ****, p <0.0001 from Mann-Whitney test.

## Tables

*Table 1 Cohort characteristics*

| | All (n=416) | Women (n=291) | Men (n=125) | P-value | FDR |
|---|---|---|---|---|---|
| Age, y | 41.01 +/- 6.35 | 40.36 +/- 6.39 | 42.52 +/- 6.01 | 0.0012 | 0.0018 |
| Baseline weight, kg | 99.69 +/- 17.14 | 95.59 +/- 15.75 | 109.37 +/- 16.44 | 1.42E-13 | 7.78E-13 |
| Baseline BMI, $kg/m^2$ | 34.76 +/- 4.87 | 34.83 +/- 4.98 | 34.59 +/- 4.63 | 0.6298 | 0.6298 |
| Baseline HOMA-IR | 2.93 +/- 2.20 | 2.73 +/- 2.25 | 3.38 +/- 2.04 | 0.0056 | 0.0069 |
| Baseline total cholesterol, mmol/L | 5.02 +/- 0.93 | 4.93 +/- 0.89 | 5.23 +/- 0.98 | 0.0043 | 0.0059 |
| Baseline LDL, mmol/L | 3.14 +/- 0.83 | 3.03 +/- 0.78 | 3.38 +/- 0.87 | 2.19E-04 | 4.81E-04 |
| Baseline HDL, mmol/L | 1.28 +/- 0.31 | 1.33 +/- 0.32 | 1.15 +/- 0.27 | 4.00E-08 | 1.47E-07 |
| Baseline waist circumference, cm | 106.89 +/- 12.75 | 103.90 +/- 12.19 | 114.10 +/- 11.13 | 1.45E-14 | 1.59E-13 |
| Percentage of weight loss during LCD | -11.06 +/- 2.74 | -10.68 +/- 2.41 | -11.90 +/- 3.22 | 0.0004 | 0.0008 |
| Percentage of weight loss at study termination | -10.79 +/- 5.97 | -10.50 +/- 5.83 | -11.59 +/- 6.32 | 0.2160 | 0.2376 |

Number corresponds to mean value +/- standard error. The t-test compares differences between men and women. BMI: Body Mass Index, FDR: False Discovery Rate, HOMA-IR: Homeostasis model assessment of insulin resistance, LCD: Low caloric diet.

# Figures

Figure 1

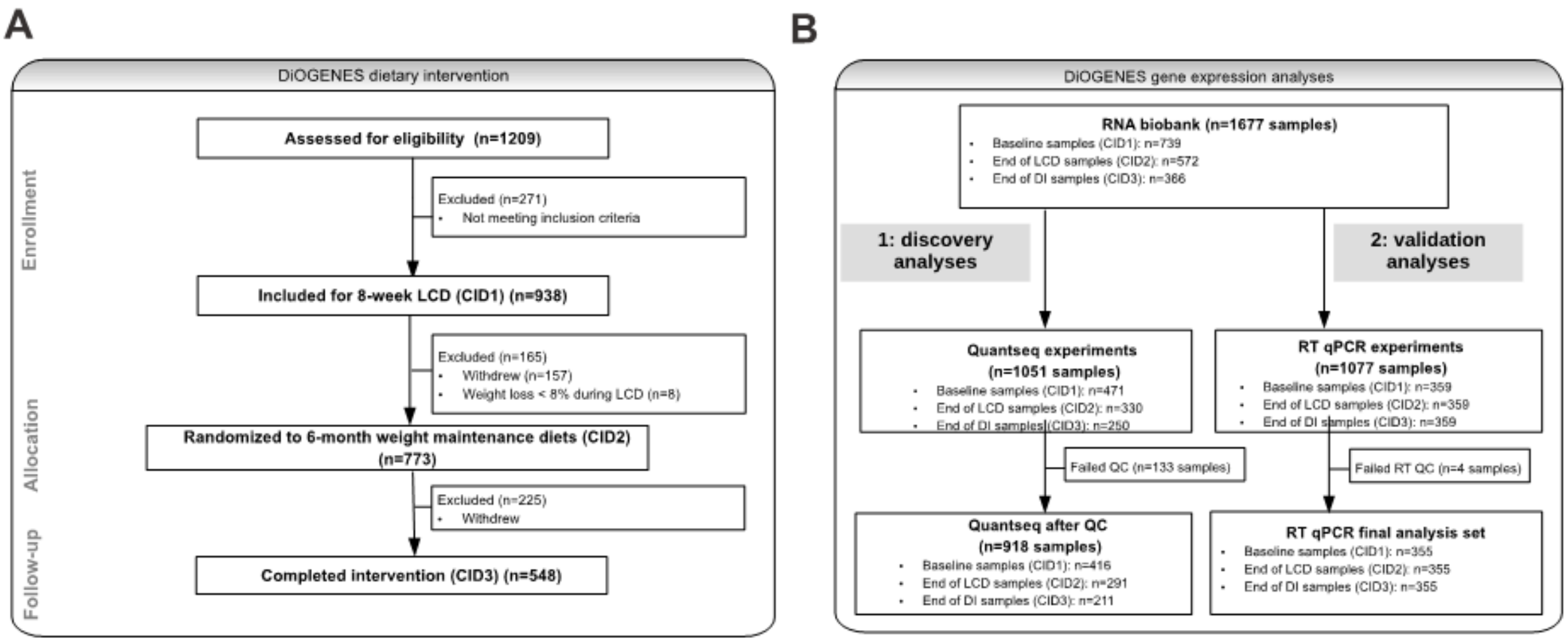

Figure 2

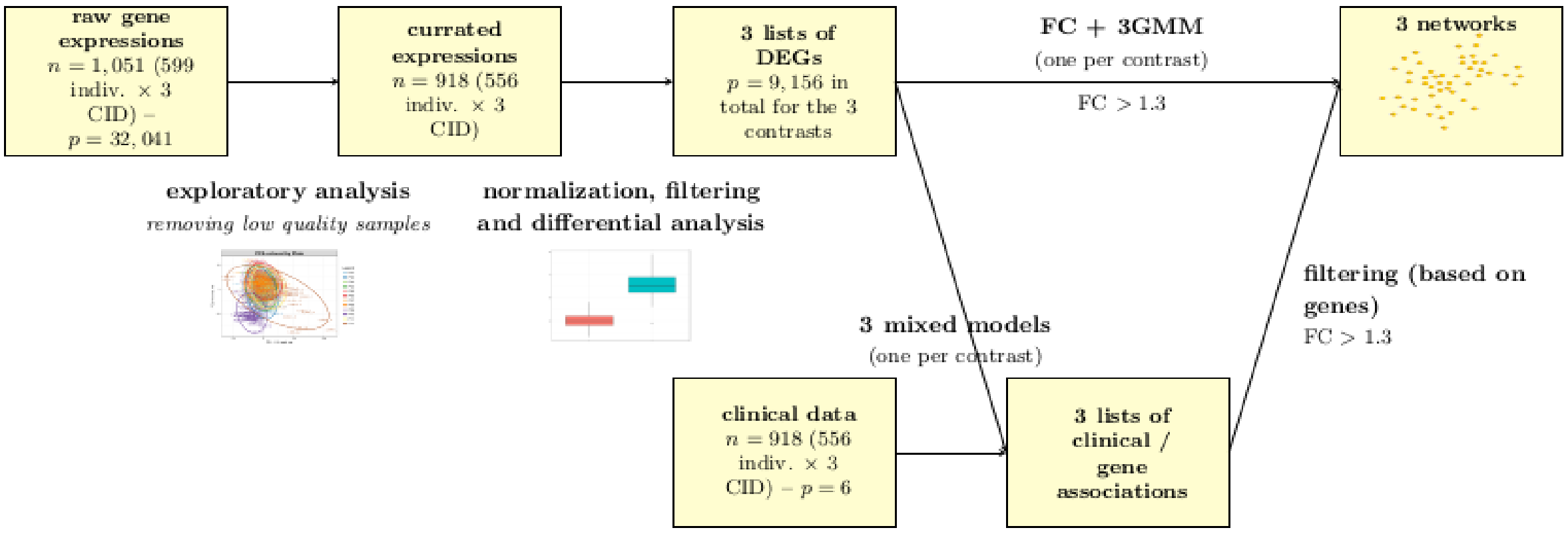

Figure 3

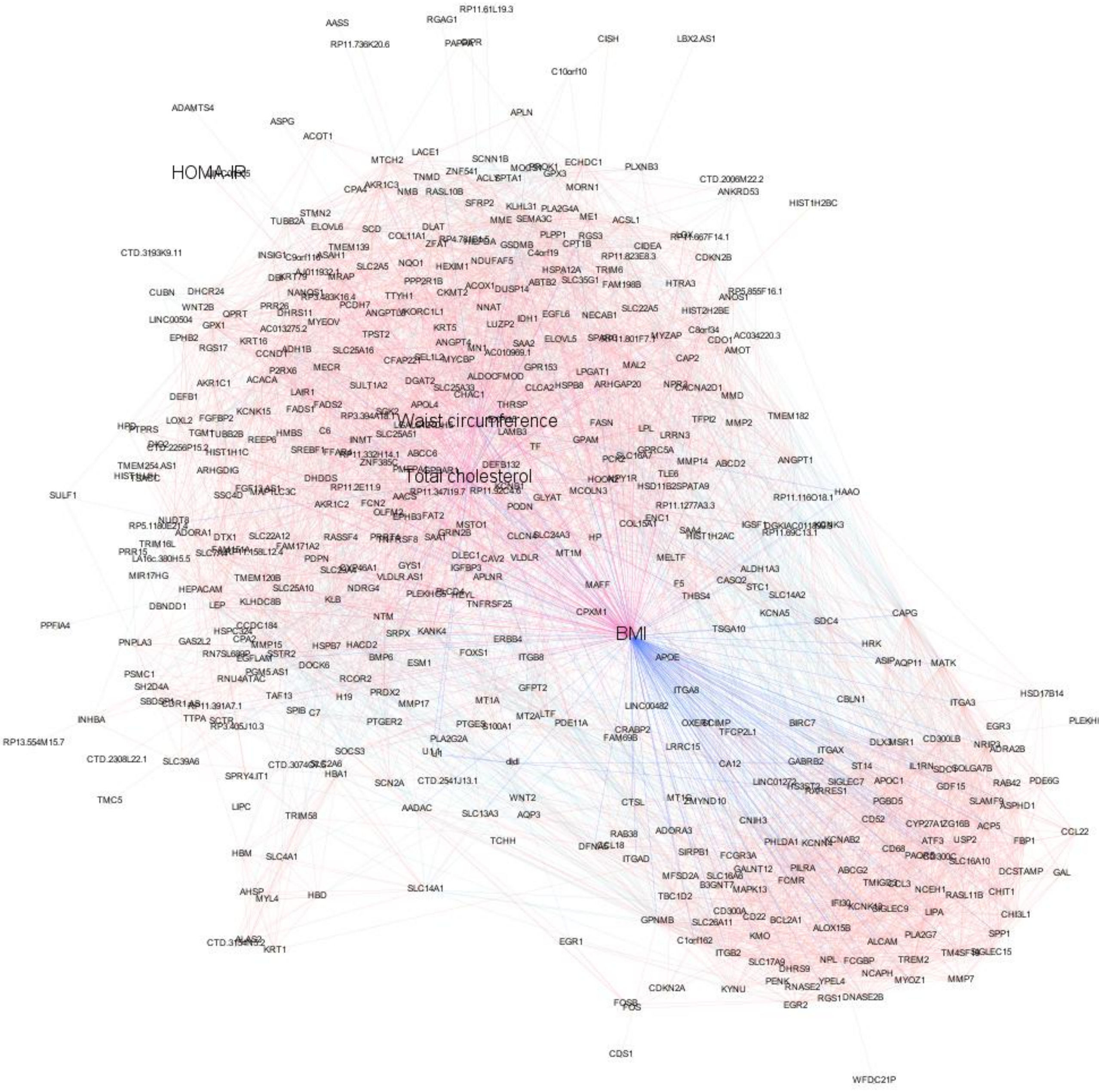

Figure 4

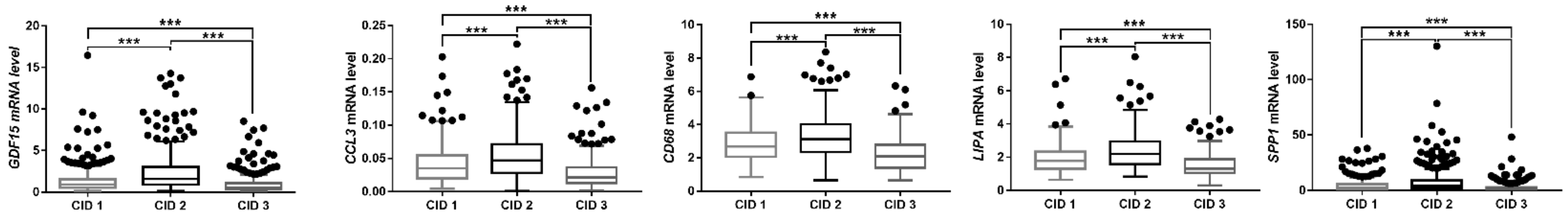

Figure 5

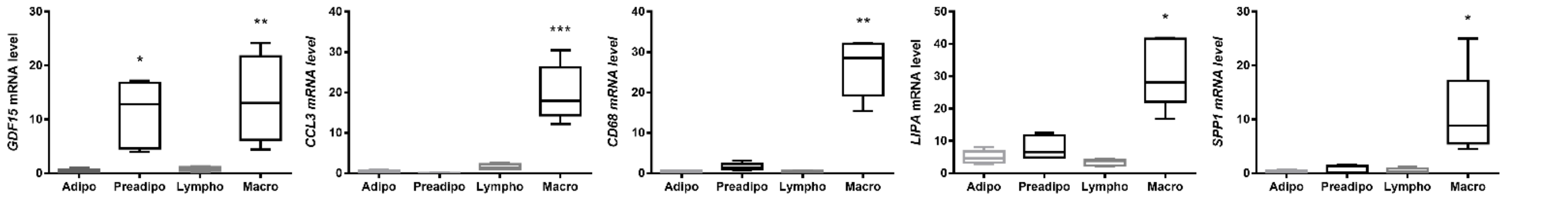

Figure 6

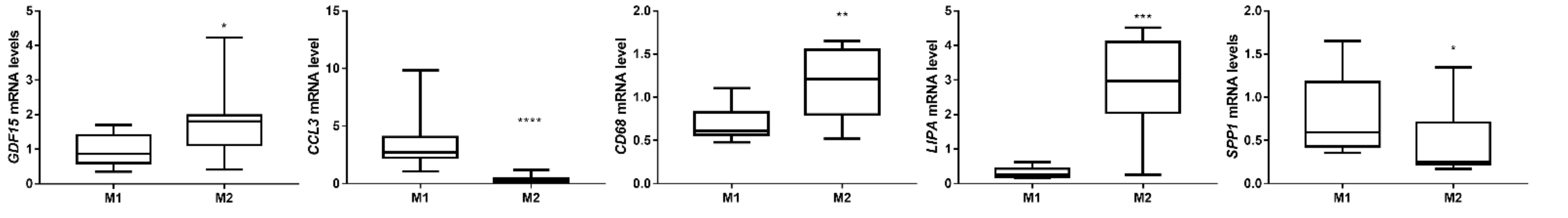